\documentclass[submission, Phys]{SciPost}

\usepackage{bm}
\usepackage{amsmath}
\usepackage{amssymb}
\usepackage{tikz}
\usetikzlibrary{calc}
\usetikzlibrary{shapes.geometric}

\begin{document}

\begin{center}
{\Large \textbf{General continuum model for twisted bilayer graphene
    and arbitrary smooth deformations}}
\end{center}

\begin{center}
L. Balents\textsuperscript{1,2*}
\end{center}

\begin{center}
{\bf 1} Kavli Institute for Theoretical Physics, University of California, Santa Barbara, CA 93106-4030
\\
{\bf 2} Canadian Institute for Advanced Research, Toronto, Ontario, Canada
\\
* balents@kitp.ucsb.edu
\end{center}

\begin{center}
\today
\end{center}


\section*{Abstract} {\bf We present a simple derivation of a continuum
  Hamiltonian for bilayer graphene with an arbitrary smooth
  lattice deformation -- technically in a fashion parametrized by displacement
  fields with small gradients.  We show that this subsumes the
  continuum model of Bistritzer and Macdonald for twisted bilayer
  graphene\cite{bistritzer2011moire} as well as many generalizations
  and extensions of it.  The derivation is carried out entirely in
  real space.  }

\vspace{10pt}
\noindent\rule{\textwidth}{1pt}
\tableofcontents\thispagestyle{fancy}
\noindent\rule{\textwidth}{1pt}
\vspace{10pt}

\section{Introduction}
\label{sec:intro}

The experimental discovery of correlated insulators and
superconductivity in magic angle Twisted Bilayer Graphene (TBG)
\cite{cao2018correlated,cao2018unconventional} has galvanized
experimental research
on these structures\cite{yankowitz2019tuning,choi19:_elect,Sharpe605,yoo19:_atomic_waals,lu2019superconductors,kerelsky19:_maxim,Kim3364}.  The bedrock for theory in these systems is the
seminal work by Bistritzer and Macdonald\cite{bistritzer2011moire} (BM),
who developed a continuum model (CM) for TBG, and by solving it, predicted
the presence of magic angles at which the bands closest to the Fermi
energy become anomalously flat.  The CM achieves several key
simplifications over microscopic lattice models.  First, it restores
periodicity, enabling the application of Bloch's theorem to TBG, which is,
strictly speaking, generally only a {\em quasi}periodic system.  Second, it
reduces the two basic dimensionless parameters of the problem -- the
ratio of interlayer to intralayer hopping, and the twist angle -- to a
single one.

The CM of BM has been the basis for most theoretical studies of TBG,
and has been elaborated and extended in various ways.  Notably, the
effect of lattice relaxation is believed to be quantitatively
important in isolating the flat bands from those further from the
Fermi energy near the first magic angle\cite{nam2017lattice}.  More
generally, non-ideal strains and inhomogeneity are increasingly
studied both theoretically and experimentally\cite{uri2019mapping,xu2019giant,PhysRevB.100.035448}.  Strong coupling of
the acoustic phonon modes of graphene to TBG has also been discussed
within the CM as a mechanism for superconductivity and linear-in-temperature resistivity in the normal state\cite{PhysRevLett.122.257002,PhysRevLett.121.257001,PhysRevB.98.241412,PhysRevB.99.165112}.

The derivation by BM  is based on an
explicit calculation of the overlap matrix elements between
quasi-momentum states in the two nearby graphene layers for rigidly
rotated sheets, carried out entirely in momentum space.  Transformed
back into real space, they obtain an intuitive form of spatially
dependent inter-layer hopping.  Later works either follow the same
procedure or simply lift the earlier result.  Further effects have
been added incrementally, e.g. external strains in
Ref.\cite{PhysRevB.100.035448}.

In the present paper, we present a path to obtain BM's model directly
in real space, which we hope has conceptual value in its economy.  To
do so, we view the twist of the bilayer as a special case of a
displacement gradient, following the classical theory of elasticity.
We show that any deformation described by an inhomogeneous
displacement field (with a small gradient) in each layer can be
treated with accuracy equal to that of the case of a uniform small
angle twist.  In this way, we obtain a model for arbitrary elastic
deformations of the sheets, which applies beyond TBG, and indeed
subsumes other generalizations of the BM model.

\section{Effective field theory and a local description}
\label{sec:effect-field-theory}

To proceed, we recall the reasoning which justifies the envelope
function approximation in semiconductors, or more generally the
applicability of continuum effective field theory.  The general
principles of the latter are that, first: the continuum description of
a perturbed system applies when the perturbations are weak and slowly
varying on the lattice scale.  Second: the continuum Hamiltonian
consists of the integral of a {\em local} Hamiltonian density which is an
analytic function of the perturbations and expressable in an expansion
in these perturbations and in small gradients.  In our case we view
the ideal graphene layer as the unperturbed system and the inter-layer
coupling and elastic deformation of the layers as perturbations.  

Consider a general elastic deformation of a single graphene layer.
The textbook description is through the displacement field
$\bm{\tilde{u}}(\bm{R})$ defined so that the location of the point
$\bm{R}$ in the undeformed solid becomes $\bm{x}$ with
\begin{equation}
  \label{eq:1}
  \bm{x} = \bm{R} + \bm{\tilde{u}}(\bm{R}).
\end{equation}
The purpose of the tilde on the displacement field will be revealed
shortly.  The deformation is locally small if $\partial_\mu
\bm{\tilde{u}} \ll 1$, and elasticity theory applies.  However, this
does not mean that $\bm{\tilde{u}}(\bm{R})$ is small, and indeed we do
not wish to make this assumption (it is invalid specifically in the
most interesting case of a layer rotation).

The case of large displacements makes the use of the
$\bm{\tilde{u}}(\bm{R})$ field problematic, because the argument
$\bm{R}$ of $\bm{\tilde{u}}$ no longer represents the location of a point in physical
space.  When we consider two layers with different large deformations, a
single  value of $\bm{R}$ corresponds to two very different physical
locations, so that locality is not manifest in these variables.  The
problem is associated with the ``Lagrangian'' coordinates adopted in
Eq.~(\ref{eq:1}), where the deformation is  associated with a point in
the undistorted lattice.  The solution of this problem is to work
instead in ``Eulerian'' coordinates, in which the displacement field $\bm{u}(\bm{x})$
is associated with an actual location in the deformed solid:
\begin{equation}
  \label{eq:2}
  \bm{x} = \bm{R} + \bm{u}(\bm{x}).
\end{equation}
This equation implicitly defines $\bm{x}(\bm{R})$ from
$\bm{u}(\bm{x})$, and while it may be less familiar to e.g. readers of
Ashcroft+Mermin, it has the distinct advantage that $\bm{u}(\bm{x})$
actually describes the deformation of the graphene sheet at position
$\bm{x}$.  Hence, the effective Hamiltonian density will be a local
functional of $\bm{u}(\bm{x})$.  It may
be helpful to point out that the Eulerian displacement field used here
is a two-dimensional analog of the phase coordinate of a charge
density wave, which describes a one-dimensional periodic structure in
a similar Eulerian fashion\cite{PhysRevB.19.3970,PhysRevB.17.535}.

\section{Construction of the continuum model}
\label{sec:constr-cont-model}

Now we are in a position to derive the CM.  It will be expressed in
terms of a continuum description of the electrons, i.e.\ a Dirac field
for each valley and layer, as well as a continuum description of the
lattice, i.e.\ an Eulerian displacement field for each layer.   There
are three physical effects  we need to
capture:
\begin{enumerate}
\item {\bf Pure geometry:} A displacement changes the physical
  positions of atoms, and a non-uniform displacement changes lengths.
  This is particularly important when the displacements of the two
  layers are different.  We must describe the physics in a
  global laboratory coordinate frame, which necessarily cannot follow
  that of both layers in that case.  
\item {\bf Intra-layer energetics:} In addition to the above purely
  geometrical effect, deforming the atoms in a layer (i.e. strain)
  changes its energetics, for example through the distance dependence
  of the hopping matrix elements.  The physical effect of strain on
  graphene has been intensively studied, and is understood to be
  represented by an ``artificial gauge
  field''\cite{suzuura2002phonons,neto2009electronic}, which is equal and opposite for
  the two graphene valleys (in order that strain preserves
  time-reversal symmetry).
\item {\bf Inter-layer coupling:} It is well-known that a small
  displacement of one layer relative to one another (a fraction of a
  lattice constant) completely modifies the inter-layer hopping
  (e.g. changing from AA to AB alignment),
  drastically changing the energy dispersion.  We
  need to keep track of this same physics {\em locally} in the bilayer
  with slowly-varying displacements.
\end{enumerate}
In the following, we will discuss each of these effects in turn, and how
they are incorporated into a continuum description.

\subsection{Transformation to global coordinates}
\label{sec:transf-glob-coord}

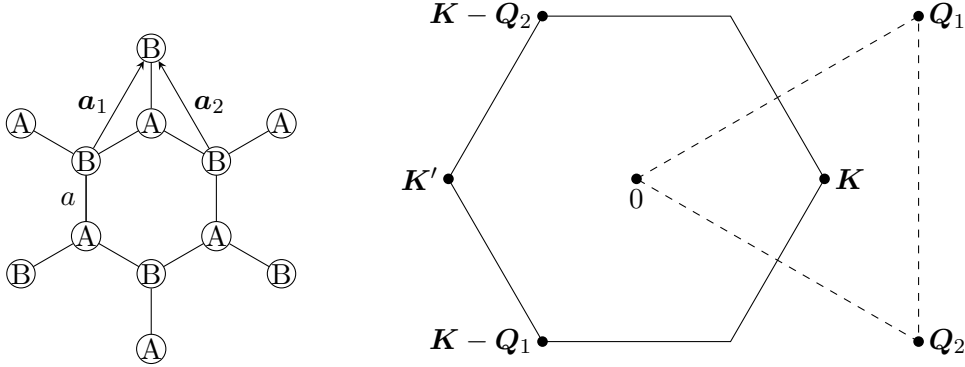
\begin{figure}[h!]
  \centering
  \begin{tikzpicture}[pt/.style={draw,circle,inner sep=0pt,fill=white}]
    \draw (0,1) node[pt](p1) {A} -- (-.866,0.5) node[pt](p2) {B} -- (-.866,-.5) node[pt](p3)
    {A} -- (0,-1) node[pt](p4) {B}-- (0.866,-.5) node[pt](p5) {A}
    -- (0.866,0.5) node[pt](p6) {B} -- cycle;
    \draw (p1) -- (0,2) node[pt](p8) {B};
    \draw (p2)--(-1.732,1) node[pt] {A};
    \draw (p6)--(1.732,1) node[pt] {A};
  \draw (p4) -- (0,-2) node[pt] {A};
    \draw (p3)--(-1.732,-1) node[pt](p7) {B};
    \draw (p5)--(1.732,-1) node[pt] {B};
       \draw (p3)--(p2) node[pos=.5,left]{$a$};
    \draw[->,>=stealth] (p2)--(p8) node[pos=.5,left]{$\bm{a}_1$};
    \draw[->,>=stealth] (p6)--(p8) node[pos=.5,right]{$\bm{a}_2$};
  \end{tikzpicture}
  \hspace{1cm}
    \begin{tikzpicture}[]
        \node[regular polygon,regular polygon sides=6, minimum size=5cm,
        draw] at (0,0) {};
        \fill [black] (0:2.5) circle (2pt) node[right] {$\bm{K}$};
        \fill [black] (240:2.5) circle (2pt) node[left]
        {$\bm{K}-\bm{Q}_1$};
        \fill [black] (120:2.5) circle (2pt) node[left] {$\bm{K}-\bm{Q}_2$};
        \fill [black] (180:2.5) circle (2pt) node[left] {$\bm{K}'$};
        \fill [black] (0,0) circle (2pt) node[below] {$0$};
        \fill [black] (30:4.33) circle (2pt) node[right] {$\bm{Q}_1$};
        \fill [black] (-30:4.33) circle (2pt) node[right]
        {$\bm{Q}_2$};
        \draw[dashed] (0,0) -- (30:4.33) -- (-30:4.33) -- cycle;
  \end{tikzpicture}
  \caption{Left panel: conventions for graphene lattice.  Two linearly
    independent Bravais lattice (translation) vectors $\bm{a}_1$,
    $\bm{a}_2$ are shown.  The distance between neighboring carbon
    atoms is $a$, as marked.   Right panel: Graphene Brillouin zone and some other
    useful wavevectors.  The wavevectors $\bm{Q}_1$ and $\bm{Q}_2$ are
    basis vectors for the reciprocal lattice.  The $\bm{K}$ point
    (also denoted $K$ point in the text) is
    the centroid of the triangle formed by the origin, $\bm{Q}_1$ and
    $\bm{Q}_2$.  The two other Brillouin zone corners
    $\bm{K}-\bm{Q}_1$ and $\bm{K}-\bm{Q}_2$ are equivalent to $\bm{K}$
    as quasimomenta, and are obtained from the latter by $C_3$
    rotations.  }
  \label{fig:honeytest}
\end{figure}

Let us imagine that the tight binding Hamiltonian in each single layer of graphene is
completely unmodified, but we associate each unit cell center,
ideally at point $\bm{R}$, with a deformed point at $\bm{x}$ described
by $\bm{u}(\bm{x})$ as in Eq.~(\ref{eq:2}).  By assumption, in the
original coordinates the physics is unchanged, so we may define
continuum Dirac fields $\psi_\pm(\bm{R})$ in the
standard way, with the subscript denoting
the valley,  via (see e.g.\ Ref.\cite{neto2009electronic} for a standard
analysis deriving the Dirac Hamiltonian and continuum fields)
\begin{equation}
  \label{eq:3}
  c(\bm{R}) \sim  \psi_+(\bm{R}) e^{i\bm{K}\cdot\bm{R}} +
  \psi_-(\bm{R}) e^{-i\bm{K}\cdot\bm{R}},
\end{equation}
where $\bm{K}=(4\pi/(3\sqrt{3}a),0)$ ($a$ is the distance between two
carbon atoms) is the zone corner momentum of one valley, see Fig.~\ref{fig:honeytest}.
We will focus on the $K$ valley, as within the continuum description
the valleys do not mix, and henceforth drop the valley index $\psi_+
\rightarrow \psi$.  Results for the other valley are readily derived
by $C_2$ symmetry.  Note that the lack of mixing is an excellent
approximation when the displacement gradients are small, because then
the slowly-varying nature of the perturbations cannot generate
sufficient momentum to mix the two valleys.  The Hamiltonian in the original coordinates has
the simple Dirac form:
\begin{equation}
  \label{eq:38}
  H_{\rm Dirac} = -iv\sum_{\mu=1,2}\int\! d^2\bm{R}\, \psi^\dagger 
  \tau^\mu \frac{\partial}{\partial R_\mu} \psi^{\vphantom\dagger},
\end{equation}
where $\tau^\mu$ are the Pauli matrices acting in the sublattice
space, and $v$ is the Dirac velocity.  
Now we must transform to the global coordinates $\bm{x}$.  We must determine
the transformation property of the Dirac field.  To obtain it, we change
coordinates in Eq.~(\ref{eq:3}) using Eq.~(\ref{eq:2}) to obtain
(again keeping only terms associated with the $K$ valley):
\begin{equation}
  \label{eq:41}
  c(\bm{x}) =\left|{\rm det}\left(
    \frac{\partial R_\mu }{\partial x_\nu}\right)\right|^{1/2} c(\bm{R}(\bm{x})) \sim\left( 1- \bm{\nabla}\cdot\bm{u}\right)^{1/2}  \psi(\bm{R})
  e^{i\bm{K}\cdot(\bm{x}-\bm{u}(\bm{x}))} \equiv \psi(\bm{x}) e^{i\bm{K}\cdot\bm{x}},
\end{equation}
where the determinant factor is the usual conformal transformation
factor for Dirac fermions, which ensures that the transformed field
$\psi(\bm{x})$ has canonical commutation relations.
This implies that
\begin{equation}
  \label{eq:42}
  \psi(\bm{x}) = \left( 1- \bm{\nabla}\cdot\bm{u}\right)^{1/2}  \psi(\bm{R}) e^{-i\bm{K}\cdot\bm{u}(\bm{x})}
  \leftrightarrow  \psi(\bm{R}) = \frac{\psi(\bm{x})}{\left( 1- \bm{\nabla}\cdot\bm{u}\right)^{1/2} } e^{i\bm{K}\cdot\bm{u}(\bm{x})}.
\end{equation}
Now we can introduce Eq.~(\ref{eq:42}) into the Dirac equation,
Eq.~(\ref{eq:38}), and change variables from $\bm{R}$ to $\bm{x}$.  We
require the field transformation and the transformation of the
integration measure and of the gradients.  The measure is
\begin{equation}
  \label{eq:43}
  d^2\bm{R}  = d^2\bm{x} \,{\rm det}\left(
    \frac{\partial R_\mu }{\partial x_\nu}\right) \approx d^2\bm{x} \left( 1- \bm{\nabla}\cdot\bm{u}\right).
\end{equation}
Note that the determinental factor in the field transformation,
Eq.~\eqref{eq:42}, cancels against the one from the integration
measure, Eq.~\eqref{eq:43} -- this is because the massless Dirac
fermion is a conformal field theory.
The gradient transforms
according to
\begin{equation}
  \label{eq:44}
   \frac{\partial}{\partial R_\mu} = \frac{\partial x_\nu}{\partial
    R_\mu} \frac{\partial}{\partial x_\nu} \approx \frac{\partial}{\partial
    x_\mu} + \frac{\partial u_\nu}{\partial x_\mu}
  \frac{\partial}{\partial x_\nu}.
\end{equation}
Both Eq.~(\ref{eq:43}) and Eq.~(\ref{eq:44}) are given to first order
in displacement gradients.  Using now
Eqs.~(\ref{eq:42},\ref{eq:43},\ref{eq:44}) in Eq.~(\ref{eq:38}) and
keeping terms to linear order in displacement gradients, we obtain
finally
\begin{equation}
  \label{eq:45}
  H_{\rm Dirac} = \int\! d^2\bm{x}\, \psi^\dagger_l \left[ -i v\left(
      \tau^\mu + \frac{\partial u_{l,\mu}}{\partial
        x_\nu}  \tau^\nu \right)\frac{\partial}{\partial x_\mu} + v
    \left(\bm{K}\cdot \partial_\mu \bm{u}_l\right)
    \tau^\mu\right] \psi^{\vphantom\dagger}_l.
\end{equation}
Here we have introduced a layer index $l$, though the individual
layers are decoupled at this point.  Note that all the changes in
Eq.~(\ref{eq:45}) from Eq.~(\ref{eq:38}) are entirely due to the
change of coordinates.  Absolutely nothing of the electron dynamics
has changed (yet).  Successive terms containing $\bm{u}_l$ in
Eq.~(\ref{eq:45}) represent: (1) a
linear transformation of spin matrices coming from the gradient
transformation;  and (2), a shift of the Dirac point,
coming from the field redefinition.   Specifically, demanding that the
energy vanish at the location of the shifted  Dirac point, we obtain
the Dirac point shift
\begin{equation}
  \label{eq:54}
  \delta\bm{K}_l = - K_\mu \bm{\nabla} u_{l,\mu} = - K \bm{\nabla} u_{l,x},
\end{equation}
to linear order in displacement gradients (here $K=|\bm{K}|$).  Owing to the lack of a
derivative action on the Dirac fields in the shift term, this is
the most important effect of lattice deformations accounted for in Eq.~(\ref{eq:45}).

\subsection{Intra-layer energetics}
\label{sec:intra-layer-strain}

Now we consider the {\em physical} effects of deforming the graphene
layers on the energetics within each layer.  This is the problem of
single layer graphene with strain, which has been extensively studied
since the early work of Ando in the context of carbon nanotubes in
2002\cite{suzuura2002phonons}.  The effect of the strain is to induce
an ``artificial gauge field'' $\mathcal{A}$ which enters the Dirac
Hamiltonian
\begin{equation}
  \label{eq:55}
  H_{\rm agf} =  \int\! d^2\bm{x}\, \psi^\dagger_l v \bm{\mathcal{A}}_l\cdot
  \bm{\tau} \psi^{\vphantom\dagger}_l.
\end{equation}
Here we can express the gauge field in terms of strain as
\begin{equation}
  \label{eq:56}
  \bm{\mathcal{A}}_l = \frac{\gamma}{a} \begin{pmatrix}
    \epsilon^l_{xx}-\epsilon^l_{yy} \\ 2 \epsilon^l_{xy}\end{pmatrix},
\end{equation}
where $\gamma$ is an order one constant proportional to $\partial \ln
t/\partial \ln a$ -- the logarithmic derivative of the hopping strength
with respect to bond length -- and $\epsilon^l$ is the strain tensor in layer $l$,
\begin{equation}
  \label{eq:4}
    \epsilon^l_{\mu\nu} = \frac{1}{2}\left( \partial_\mu u_{l,\nu} +
    \partial_\nu u_{l,\mu}\right).
\end{equation}
The artificial gauge field here is that of the $K$ valley.  Of of
opposite sign appears for the $K'$
valley, thereby maintaining time-reversal symmetry.

Comparing to Eq.~(\ref{eq:45}), we see that the artificial gauge field
is of the same magnitude as the $K$-point shift, because both
contributions are proportional to a single gradient of displacement,
with a coefficient of order  the Dirac velocity divided by the
microscopic lattice spacing (note $K \propto 1/a$).  So, for a generic
strained layer, this is an important effect.  However, for a rigid
rotation, the strain tensor vanishes and so does the artificial gauge
field.  Hence, if the rotation is not too inhomogeneous, we can neglect
the strain effect in this section completely.  

\subsection{Interlayer tunneling}
\label{sec:interlayer-tunneling}

What remains is to consider the coupling between layers. We will
assume that the main coupling mechanism, at the single particle level,
is  tunneling between layers, a standard assumption, made e.g. by BM\cite{bistritzer2011moire}.  
We proceed to determine the form of the tunneling contribution using a
symmetry analysis.  In general, the tunneling contribution to the
Hamiltonian can be written as
\begin{equation}
  \label{eq:57}
  H_{\rm tun} = \int\! d^2\bm{x}\, \psi_2^\dagger \, {\sf
    T}(\bm{u}_1-\bm{u}_2)\,
  \psi_1^{\vphantom\dagger} + {\rm h.c.},
\end{equation}
where ${\sf T}(\bm{u})$ is a 2$\times$2 matrix in the sublattice space, which
depends upon the {\em difference} $\bm{u}\equiv \bm{u}_1-\bm{u}_2$ of the displacement fields.  In
writing Eq.~(\ref{eq:57}) we have used several principles.  First, we have
assumed spin-independent tunneling, which is justified by the
negligible spin-orbit coupling in graphene.  Next, we have assumed that the tunneling
term is local in the continuum model, as is the standard assumption in
effective field theory.  Furthermore, we neglected any dependence upon the
gradients of the displacements.  The latter is certainly present, but
sub-dominant.  By the standard analyticity assumption of effective
field theory, $H_{\rm tun}$ is an analytic function of the gradients,
and because they are assumed small, $H_{\rm tun}$ is dominated by its zeroth
order term.  To be a bit more precise, we are assuming there are two
small parameters: the tunneling strength and the size of the
displacement gradients.  We are keeping terms in the Hamiltonian
linear in these parameters, but not higher order ones.

Notably, unlike the effects in the previous two subsections, the
tunneling energy in Eq.~(\ref{eq:57}) depends explicitly on the
displacements themselves (and not only on their gradients).  The
single layer terms could not be functions of the displacements
directly, because the energy of a single layer must be independent of
a uniform displacement of that layer.  The inter-layer energy can and
does depend upon their relative displacement.  However, it must be
independent of a uniform and simultaneous translation of both layers,
which is why it is a function only of the difference
$\bm{u}_1-\bm{u}_2$.

To further constrain the tunneling matrix, we take into account
remaining symmetries.  First, we have a remaining {\em discrete}
translational symmetry: the energy must be independent of a constant
displacement of one layer (relative to the other) by a Bravais lattice
vector. This implies
the condition
\begin{equation}
  \label{eq:58}
  {\sf T}(\bm{u}+\bm{a}) = {\sf T}(\bm{u}), \qquad \bm{a}\in \textrm{B.L.}
\end{equation}
Consequently, we can represent the tunneling matrix in a Fourier
series
\begin{equation}
  \label{eq:59}
  {\sf T}(\bm{u}) = \sum_{\bm{Q}\in R.L.} {\sf T}_{\bm{Q}}e^{i\bm{Q}\cdot\bm{u}},
\end{equation}
where the sum is over reciprocal lattice vectors, and ${\sf
  T}_{\bm{Q}}$ are Fourier coefficients.

There are a few remaining symmetries to use.  We consider here only
those symmetries which preserve the valley -- others serve to
determine the Hamiltonian for the other valley from this one.  In
applying these symmetries, we take the origin of the un-displaced
lattice to lie at the center of a hexagon, and both layers to be aligned
directly atop one another.  The symmetry operations are: 
\begin{enumerate}
\item $C_2$T: The combination of time-reversal symmetry with a
  two-fold rotation through the hexagon center (of the
  un-displaced lattice).
\item $R_y$: $y\rightarrow -y$ reflection through a plane cutting mid-way
  through a carbon-carbon bond,
\item $C_3$ rotation about the center of a hexagon (of the
  un-displaced lattice).
\end{enumerate}
Note that the operations are symmetries only when both the electrons
{\em and} the displacements are transformed, because any given
configuration of displacements typically is not invariant under the operations.

To use these symmetries, we need to understand how they act on the
fields $\psi_l(\bm{x})$ and $\bm{u}_l(\bm{x})$.  The transformations
of $\bm{u}_l$ are same as those of the coordinates, as it is a vector,
and those of $\psi$ can be inferred by demanding the invariance of
Eq.~(\ref{eq:45}).  Putting this together, one finds that under $C_2$T,
\begin{equation}
  \label{eq:60}
  C_2T: \qquad \psi_l(\bm{x}) \rightarrow \mathcal{C} \tau^x
  \psi_l(-\bm{x}), \qquad \bm{u}_l(\bm{x}) \rightarrow - \bm{u}_l(-\bm{x}).
\end{equation}
Here $\mathcal{C}$ is the complex conjugation operator.
Under the $y$ reflection,
\begin{equation}
  \label{eq:61}
  R_y: \qquad \psi_l(x,y) \rightarrow \tau^x
  \psi_l(x,-y), \qquad \bm{u}_l(x,y) \rightarrow   {\sf O}_2\bm{u}_l(x,-y),
\end{equation}
where ${\sf O}_2 = \begin{pmatrix} 1 &0\\ 0 &-1\end{pmatrix}$.
Finally, the $C_3$ rotation is the most non-trivial:
\begin{align}
  \label{eq:62}
  C_3: \qquad &\psi_l(\bm{x}) \rightarrow e^{i\bm{Q}_1\cdot\bm{u}_l(\bm{x}')}
  e^{2\pi i \tau^z/3} \psi_l(\bm{x}'), \qquad \bm{u}_l(\bm{x})
  \rightarrow {\sf O}_3^{-1} \bm{u}_l(\bm{x}'), \nonumber \\
  & \bm{x}' = {\sf O}_3 \,\bm{x},
\end{align}
with
\begin{equation}
  \label{eq:63}
 {\sf O}_3 = \begin{pmatrix} -\frac{1}{2} &
    \frac{\sqrt{3}}{2} \\ -\frac{\sqrt{3}}{2} & -\frac{1}{2} \end{pmatrix}.
\end{equation}
Here $\bm{Q}_1$ is one of the shortest graphene reciprocal lattice
vectors (see Fig.~\ref{fig:honeytest} and Eq.~(\ref{eq:73})), and ${\sf O}_3$
is a $C_3$ rotation matrix.  The
$e^{2\pi i \tau^z/3}$ factor in front of the transformed $\psi$ in Eq.~(\ref{eq:62}) ensures
that as the coordinate $\bm{x}$ is rotated, the sublattice Dirac
matrices follow.  The exponential of the displacement field compensates for the rotation of the $\bm{K}$ point (at the
``eastern'' corner of the Brillouin zone) to its
equivalent point (at the ``southwestern'' corner), which is separated
from its original position by $\bm{Q}_1$.

With these transformations in hand, we can now proceed to constrain
the tunneling matrix coefficients ${\sf T}_{\bm Q}$.    We assume that the
tunneling has a non-zero spatial average, which requires the
$\bm{Q}=\bm{0}$ term to be non-zero.  The matrix ${\sf T}_{\bm{0}}$ is constrained by the $C_2T$ and $R_y$ symmetries.  Specifically,
they imply the two conditions:
\begin{align}
  \label{eq:64}
  C_2T: \qquad {\sf T}_0 = \tau^x {\sf T}^*_0\tau^x , \nonumber\\
  R_y: \qquad {\sf T}_0 = \tau^x {\sf T}_0\tau^x.
\end{align}
The most general two-dimensional matrix satisfying Eqs.~(\ref{eq:64}) is
\begin{equation}
  \label{eq:65}
  {\sf T}_0 = u \mathbb{I} + w \tau^x = u\mathbb{I} + w(\tau^+ + \tau^-),
\end{equation}
where $u$ and $w$ are {\em real} parameters.  We will determine their
physical significance shortly.

Now we address the other $\bm{Q}$.  We expect that tunneling decays
for large $|\bm{Q}|$.  Therefore, we will derive a {\em minimal}
description which assumes the only other non-zero ${\sf T}_{\bm{Q}}$
are those which are required to be non-zero to be consistent with
symmetry.  This is accomplished by applying the condition of $C_3$
invariance, which relates other Fourier coefficients of ${\sf T}$ to
${\sf T}_{\bm{0}}$.  Invariance under $C_3$ (as given in Eq.~(\ref{eq:62})) of the
full tunneling term in Eq.~(\ref{eq:57}) gives
\begin{equation}
  \label{eq:66}
  {\sf T}(\bm{u}) = e^{-i \bm{Q}_1\cdot\bm{u}} e^{-2\pi i \tau^z/3}
  {\sf T}({\sf O}_3^{-1}\bm{u}) e^{2\pi i \tau^z/3}.
\end{equation}
Now we put the Fourier expansion of Eq.~(\ref{eq:59}) into
Eq.~(\ref{eq:66}), and equate coefficients on both sides.  This
implies
\begin{equation}
  \label{eq:67}
  {\sf T}_{\bm{Q}'} = e^{-2\pi i \tau^z/3}
  {\sf T}_{\bm{Q}} e^{2\pi i \tau^z/3}, \qquad \bm{Q}'= {\sf O}_3 \bm{Q}-\bm{Q}_1.
\end{equation}
Eq.~(\ref{eq:67}), when applied on an initial ${\sf T}_{\bm{Q}}$ on the
right hand side, generates  another ${\sf T}_{\bm{Q}'}$ on the left,
with $\bm{Q}'={\sf O}_3 \bm{Q}-\bm{Q}_1$.  We iterate this relation starting with
$\bm{Q}=\bm{0}$ and generate  two further Fourier coefficients before the iteration closes.

The result is
\begin{equation}
  \label{eq:72}
  {\sf T}_j \equiv T_{-\bm{Q}_j} = u \mathbb{I} + w\left(
    \bar{\zeta}^j \tau^++\zeta^j \tau^-\right), \qquad j=0,1,2,
\end{equation}
where
\begin{equation}
  \label{eq:69}
  \zeta = e^{2\pi i/3}, \qquad \bar{\zeta} = \zeta^* = \frac{1}{\zeta}
  = e^{-2\pi i/3}.
\end{equation}
and
\begin{equation}
  \label{eq:73}
  \bm{Q}_0 = 0, \qquad \bm{Q}_1 = \frac{4\pi}{3a} \begin{pmatrix} \sqrt{3}/2 \\
    1/2\end{pmatrix}, \qquad   \bm{Q}_2 = \frac{4\pi}{3a} \begin{pmatrix} \sqrt{3}/2 \\
    -1/2\end{pmatrix}.
\end{equation}
The three wavevectors $\bm{Q}_j$, $j=0,1,2$ form an equilateral
triangle of reciprocal lattice points whose centroid is the $\bm{K}$
point (see Fig.~\ref{fig:honeytest}).

Now we can assign physical significance to $u$ and $w$ by considering some
special cases.   Suppose we take $\bm{u}_l=\bm{0}$, which corresponds to an
AA bilayer.  We have
\begin{equation}
  \label{eq:79}
  {\sf T}(\bm{u}) = \sum_j {\sf T}_j e^{-i\bm{Q}_j \cdot \bm{u}}
  \xrightarrow[{\bm{u}=\bm{0}}]{} \sum_j  {\sf T}_j = 3 u \mathbb{I}. 
\end{equation}
In the final result, we used the fact that
$\sum_j \zeta^j = \sum_j \bar{\zeta}^j=0$.  Eq.~(\ref{eq:79}) agrees with the
simple physical expectation that we just have diagonal sublattice hopping
$t'$ in this configuration.  Hence we conclude that
\begin{equation}
  \label{eq:80}
  u = \frac{t'_{AA}}{3},
\end{equation}
where $t'_{AA}$ is the interlayer hopping for an AA region.  Now
consider $\bm{u}=(0,a)$, which corresponds to AB stacking.  We see
from Eq.~\eqref{eq:73} that $e^{-i \bm{Q}_j \cdot \bm{u}} =
\bar{\zeta}^j$.  Then we have
\begin{equation}
  \label{eq:81}
    {\sf T}(\bm{u}) = \sum_j {\sf T}_j e^{-i\bm{Q}_j \cdot \bm{u}}
  \xrightarrow[{\bm{u}=(0,a)}]{}\sum_j  {\sf T}_j \bar\zeta^j= 3 w \tau^-.
\end{equation}
This is again consistent with expectations: interlayer hopping happens only
from sublattice B on layer 1 to sublattice A on layer 2, or vice versa.  Hence we
see that
\begin{equation}
  \label{eq:82}
  w = \frac{t'_{AB}}{3}.
\end{equation}
We have allowed for the inter-layer hopping $t'_{AB}$ to be different
in the AB regions from that in the AA ones, though this difference is
expected to be small.

We have now fully determined all the terms in the continuum
description, by combining Eq.~(\ref{eq:45}), Eq.~(\ref{eq:55}), and
Eq.~(\ref{eq:57}), which are now fully specified.  The result is given
explicitly for clarity in Eq.~(\ref{eq:83}) of the Conclusion.

\subsection{Application to a rigid twist}
\label{sec:appl-rigid-twist}

To connect to the continuum model of BM, we specialize to a rigid
twist, and evaluate the general result for the case
\begin{equation}
  \label{eq:84}
  \bm{u}_1 = -\bm{u}_2 = \frac{\theta}{2} \bm{\hat{z}}\times \bm{x}.
\end{equation}
For such a pure rotation, the strain-induced gauge field vanishes,
$\bm{\mathcal{A}}_l=\bm{0}$, and there is zero compression so 
$\bm{\nabla}\cdot \bm{u}_l=0$ as well.  The $K$-point shift term becomes
\begin{equation}
  \label{eq:85}
  \bm{K}\cdot \partial_\mu \bm{u}_1 = - \bm{K}\cdot \partial_\mu
  \bm{u}_2 = \mp \frac{\theta K}{2} \bm{\hat{y}} \equiv
 \mp \frac{k_\theta}{2} \bm{\hat{y}},
\end{equation}
where $k_\theta = \theta K$.  The displacement
gradient itself is
\begin{equation}
  \label{eq:86}
  \frac{\partial u^{1}_\mu}{\partial x_\nu}= - \frac{\partial u^{2}_\mu}{\partial
    x_\nu} = -\frac{\theta}{2} \epsilon_{\mu\nu}.
\end{equation}
Most interestingly, the term in the exponential in the second line of
Eq.~\eqref{eq:83} becomes
\begin{equation}
  \label{eq:87}
  \bm{Q}_j \cdot \left(\bm{u}_1-\bm{u}_2\right) = \theta \bm{Q}_j
  \cdot \bm{\hat{z}}\times \bm{x} = -  \theta \bm{\hat{z}}\times
    \bm{Q}_j\cdot \bm{x}.
\end{equation}
We see that this immediately produces the wavevectors
\begin{equation}
  \label{eq:90}
  \bm{q}_j =  -\theta\bm{\hat{z}}\times \bm{Q}_j.
\end{equation}
The two non-zero vectors $\bm{q}_1,\bm{q}_2$ are two basis vectors of
the reciprocal lattice of the moir\'e pattern!  Putting this all
together, we can write the Hamiltonian for the rigid twist as
\begin{align}
  \label{eq:92}
  H_{K,\theta} =  \int\! d^2\bm{x} \Bigg\{ &\psi_1^\dagger \left[ -iv
               \bm{\tau}(\tfrac{\theta}{2}) \cdot \bm{\nabla} -
               \frac{vk_\theta}{2}
               \tau^y\right]\psi_1^{\vphantom\dagger} +  \psi_2^\dagger \left[ -iv
               \bm{\tau}(-\tfrac{\theta}{2}) \cdot \bm{\nabla} +
               \frac{vk_\theta}{2}
               \tau^y\right]\psi_2^{\vphantom\dagger} \nonumber\\
               & + \sum_j \left[ e^{-i\bm{q}_j \cdot\bm{x}} \psi_2^\dagger
    \, {\sf T}_j \psi_1^{\vphantom\dagger} +{\rm h.c.}\right] \Bigg\}.
\end{align}
Here we defined rotated Pauli matrices
\begin{equation}
  \label{eq:93}
\tau^\mu(\theta) = \tau^\mu - \theta \epsilon_{\mu\nu} \tau^\nu.
\end{equation}
The result is in perfect agreement with BM.

\section{Conclusion}
\label{sec:conclusion}

We provided a simple real space derivation of a full continuum model
for bilayer graphene in the presence of small displacement
gradients.  Combining all three contributions discussed in
Sec.~\ref{sec:constr-cont-model} gives the full continuum band
Hamiltonian for {\em arbitrary} small displacement gradients for the
$K$ valley.  It is
\begin{align}
  \label{eq:83}
  H_K[\bm{u}_1,\bm{u}_2] = \int\! d^2\bm{x} \Bigg\{  & \sum_l \psi^\dagger_l \left[ -i v\left(
      \tau^\mu +\frac{\partial u^{l}_\mu}{\partial
        x_\nu}  \tau^\nu\right)\frac{\partial}{\partial x_\mu} + v
    \left(\bm{K}\cdot \partial_\mu \bm{u}_l+\bm{\mathcal{A}}_l\right)
  \tau^\mu\right] \psi^{\vphantom\dagger}_l \nonumber \\
  & + \sum_j \left[ e^{-i\bm{Q}_j \cdot (\bm{u}_1-\bm{u}_2)} \psi_2^\dagger
    \, {\sf T}_j \psi_1^{\vphantom\dagger} +{\rm h.c.}\right] \Bigg\}.
\end{align}
We emphasize that Eq.~\eqref{eq:83} is derived under the assumption
that $\partial_\mu \bm{u}_l \ll 1$ and $t' \ll t$, but otherwise
$\bm{u}_l$ may have {\em arbitrary} spatial dependence.  As shown in
Sec.~\ref{sec:appl-rigid-twist} it exactly recovers the BM model for
the case of a rigid twist.  Eq.~\eqref{eq:83} goes far beyond this to
enable treatment of not only twisting, but also uniform and non-uniform strains.  It
also ought to be sufficient to understand the coupling of low energy
phonons (those derived from the acoustic modes of the original
graphene layers) to the bilayer -- one needs only to add dynamics to the
displacement fields.  Finally, Eq.~(\ref{eq:83}) indeed is capable of
describing all three of these effects {\em together}, simply by taking
\begin{equation}
  \label{eq:5}
  \bm{u}_l = (3-2l)\frac{\theta}{2} \bm{\hat{z}}\times \bm{x} +
  \bm{\hat{u}}_l, \qquad l=1,2,
\end{equation}
where $\bm{\hat{u}}_l$ represents the strain and/or phonons.  For
example, this is a natural point of departure to discuss the effects of
twist angle inhomogeneity in twisted bilayer graphene, as there is no
need to assume random strains are small compared to the displacement
gradient comprising the twist.  Hopefully this formulation will have
pedagogical value and find useful
applications.

\section*{Acknowledgements}

I thank the organizers and students of the EPICQUR summer school for
stimulating me to complete this derivation, and for Lucile Savary for
helpful advice on the manuscript.  This work was supported
by the DOE, Office of Science, Basic Energy Sciences under Award
No. DE-FG02-08ER46524.

\bibliography{tbg.bib}

\begin{thebibliography}{10}
\providecommand{\url}[1]{\texttt{#1}}
\providecommand{\urlprefix}{URL }
\expandafter\ifx\csname urlstyle\endcsname\relax
  \providecommand{\doi}[1]{doi:\discretionary{}{}{}#1}\else
  \providecommand{\doi}{doi:\discretionary{}{}{}\begingroup
  \urlstyle{rm}\Url}\fi
\providecommand{\eprint}[2][]{\url{#2}}

\bibitem{bistritzer2011moire}
R.~Bistritzer and A.~H. MacDonald,
\newblock \emph{Moir{\'e} bands in twisted double-layer graphene},
\newblock Proceedings of the National Academy of Sciences \textbf{108}(30),
  12233 (2011),
\newblock \doi{10.1073/pnas.1108174108}.

\bibitem{cao2018correlated}
Y.~Cao, V.~Fatemi, A.~Demir, S.~Fang, S.~L. Tomarken, J.~Y. Luo, J.~D.
  Sanchez-Yamagishi, K.~Watanabe, T.~Taniguchi, E.~Kaxiras \emph{et~al.},
\newblock \emph{Correlated insulator behaviour at half-filling in magic-angle
  graphene superlattices},
\newblock Nature \textbf{556}(7699), 80 (2018),
\newblock \doi{10.1038/nature26154}.

\bibitem{cao2018unconventional}
Y.~Cao, V.~Fatemi, S.~Fang, K.~Watanabe, T.~Taniguchi, E.~Kaxiras and
  P.~Jarillo-Herrero,
\newblock \emph{Unconventional superconductivity in magic-angle graphene
  superlattices},
\newblock Nature \textbf{556}(7699), 43 (2018),
\newblock \doi{10.1038/nature26160}.

\bibitem{yankowitz2019tuning}
M.~Yankowitz, S.~Chen, H.~Polshyn, Y.~Zhang, K.~Watanabe, T.~Taniguchi,
  D.~Graf, A.~F. Young and C.~R. Dean,
\newblock \emph{Tuning superconductivity in twisted bilayer graphene},
\newblock Science \textbf{363}(6431), 1059 (2019),
\newblock \doi{10.1126/science.aav1910}.

\bibitem{choi19:_elect}
Y.~Choi, J.~Kemmer, Y.~Peng, A.~Thomson, H.~Arora, R.~Polski, Y.~Zhang, H.~Ren,
  J.~Alicea, G.~Refael, F.~von Oppen, K.~Watanabe \emph{et~al.},
\newblock \emph{Electronic correlations in twisted bilayer graphene near the
  magic angle},
\newblock Nature Physics  (2019),
\newblock \doi{10.1038/s41567-019-0606-5}.

\bibitem{Sharpe605}
A.~L. Sharpe, E.~J. Fox, A.~W. Barnard, J.~Finney, K.~Watanabe, T.~Taniguchi,
  M.~A. Kastner and D.~Goldhaber-Gordon,
\newblock \emph{Emergent ferromagnetism near three-quarters filling in twisted
  bilayer graphene},
\newblock Science \textbf{365}(6453), 605 (2019),
\newblock \doi{10.1126/science.aaw3780},
\newblock
  \eprint{https://science.sciencemag.org/content/365/6453/605.full.pdf}.

\bibitem{yoo19:_atomic_waals}
H.~Yoo, R.~Engelke, S.~Carr, S.~Fang, K.~Zhang, P.~Cazeaux, S.~H. Sung,
  R.~Hovden, A.~W. Tsen, T.~Taniguchi, K.~Watanabe, G.-C. Yi \emph{et~al.},
\newblock \emph{Atomic and electronic reconstruction at the van der waals
  interface in twisted bilayer graphene},
\newblock Nature Materials \textbf{18}(5), 448 (2019),
\newblock \doi{10.1038/s41563-019-0346-z}.

\bibitem{lu2019superconductors}
X.~{Lu}, P.~{Stepanov}, W.~{Yang}, M.~{Xie}, M.~A. {Aamir}, I.~{Das},
  C.~{Urgell}, K.~{Watanabe}, T.~{Taniguchi}, G.~{Zhang}, A.~{Bachtold}, A.~H.
  {MacDonald} \emph{et~al.},
\newblock \emph{Superconductors, orbital magnets, and correlated states in
  magic angle bilayer graphene},
\newblock arXiv preprint arXiv:1903.06513  (2019).

\bibitem{kerelsky19:_maxim}
A.~Kerelsky, L.~J. McGilly, D.~M. Kennes, L.~Xian, M.~Yankowitz, S.~Chen,
  K.~Watanabe, T.~Taniguchi, J.~Hone, C.~Dean, A.~Rubio and A.~N. Pasupathy,
\newblock \emph{Maximized electron interactions at the magic angle in twisted
  bilayer graphene},
\newblock Nature \textbf{572}(7767), 95 (2019),
\newblock \doi{10.1038/s41586-019-1431-9}.

\bibitem{Kim3364}
K.~Kim, A.~DaSilva, S.~Huang, B.~Fallahazad, S.~Larentis, T.~Taniguchi,
  K.~Watanabe, B.~J. LeRoy, A.~H. MacDonald and E.~Tutuc,
\newblock \emph{Tunable moir{\'e} bands and strong correlations in
  small-twist-angle bilayer graphene},
\newblock Proceedings of the National Academy of Sciences \textbf{114}(13),
  3364 (2017),
\newblock \doi{10.1073/pnas.1620140114},
\newblock \eprint{https://www.pnas.org/content/114/13/3364.full.pdf}.

\bibitem{nam2017lattice}
N.~N. Nam and M.~Koshino,
\newblock \emph{Lattice relaxation and energy band modulation in twisted
  bilayer graphene},
\newblock Physical Review B \textbf{96}(7), 075311 (2017),
\newblock \doi{10.1103/PhysRevB.96.075311}.

\bibitem{uri2019mapping}
A.~Uri, S.~Grover, Y.~Cao, J.~Crosse, K.~Bagani, D.~Rodan-Legrain,
  Y.~Myasoedov, K.~Watanabe, T.~Taniguchi, P.~Moon \emph{et~al.},
\newblock \emph{Mapping the twist angle and unconventional landau levels in
  magic angle graphene},
\newblock arXiv preprint arXiv:1908.04595  (2019).

\bibitem{xu2019giant}
S.~Xu, A.~Berdyugin, P.~Kumaravadivel, F.~Guinea, R.~K. Kumar, D.~Bandurin,
  S.~Morozov, W.~Kuang, B.~Tsim, S.~Liu \emph{et~al.},
\newblock \emph{Giant oscillations in a triangular network of one-dimensional
  states in marginally twisted graphene},
\newblock arXiv preprint arXiv:1905.12984  (2019).

\bibitem{PhysRevB.100.035448}
Z.~Bi, N.~F.~Q. Yuan and L.~Fu,
\newblock \emph{Designing flat bands by strain},
\newblock Phys. Rev. B \textbf{100}, 035448 (2019),
\newblock \doi{10.1103/PhysRevB.100.035448}.

\bibitem{PhysRevLett.122.257002}
B.~Lian, Z.~Wang and B.~A. Bernevig,
\newblock \emph{Twisted bilayer graphene: A phonon-driven superconductor},
\newblock Phys. Rev. Lett. \textbf{122}, 257002 (2019),
\newblock \doi{10.1103/PhysRevLett.122.257002}.

\bibitem{PhysRevLett.121.257001}
F.~Wu, A.~H. MacDonald and I.~Martin,
\newblock \emph{Theory of phonon-mediated superconductivity in twisted bilayer
  graphene},
\newblock Phys. Rev. Lett. \textbf{121}, 257001 (2018),
\newblock \doi{10.1103/PhysRevLett.121.257001}.

\bibitem{PhysRevB.98.241412}
Y.~W. Choi and H.~J. Choi,
\newblock \emph{Strong electron-phonon coupling, electron-hole asymmetry, and
  nonadiabaticity in magic-angle twisted bilayer graphene},
\newblock Phys. Rev. B \textbf{98}, 241412 (2018),
\newblock \doi{10.1103/PhysRevB.98.241412}.

\bibitem{PhysRevB.99.165112}
F.~Wu, E.~Hwang and S.~Das~Sarma,
\newblock \emph{Phonon-induced giant linear-in-$t$ resistivity in magic angle
  twisted bilayer graphene: Ordinary strangeness and exotic superconductivity},
\newblock Phys. Rev. B \textbf{99}, 165112 (2019),
\newblock \doi{10.1103/PhysRevB.99.165112}.

\bibitem{PhysRevB.19.3970}
P.~A. Lee and T.~M. Rice,
\newblock \emph{Electric field depinning of charge density waves},
\newblock Phys. Rev. B \textbf{19}, 3970 (1979),
\newblock \doi{10.1103/PhysRevB.19.3970}.

\bibitem{PhysRevB.17.535}
H.~Fukuyama and P.~A. Lee,
\newblock \emph{Dynamics of the charge-density wave. i. impurity pinning in a
  single chain},
\newblock Phys. Rev. B \textbf{17}, 535 (1978),
\newblock \doi{10.1103/PhysRevB.17.535}.

\bibitem{suzuura2002phonons}
H.~Suzuura and T.~Ando,
\newblock \emph{Phonons and electron-phonon scattering in carbon nanotubes},
\newblock Physical review B \textbf{65}(23), 235412 (2002),
\newblock \doi{10.1103/PhysRevB.65.235412}.

\bibitem{neto2009electronic}
A.~C. Neto, F.~Guinea, N.~M. Peres, K.~S. Novoselov and A.~K. Geim,
\newblock \emph{The electronic properties of graphene},
\newblock Reviews of modern physics \textbf{81}(1), 109 (2009),
\newblock \doi{10.1103/RevModPhys.81.109}.

\end{thebibliography}

\nolinenumbers

\end{document}